\documentstyle[11pt,newpasp,epsf]{article}

\def\ha{H$\alpha$}
\def\hr{H$\alpha$--{\scr R}}
\def\u{U$_{336}$}
\def\b{B$_{439}$}
\def\v{V$_{555}$}
\def\r{R$_{675}$}
\def\i{I$_{814}$}
\def\ub{U$_{336}-$B$_{439}$}
\def\vi{V$_{555}-$I$_{814}$}
\def\hr{H$\alpha-$R$_{675}$}

\def\et{{\it et al.}}
\def\asec{$''$}
\def\amin{$'$}
\def\deg{$^{\rm o}$}

\def\secspt{$\buildrel{\rm s}\over .$}
\def\asecspt{$\buildrel{\prime\prime}\over .$}
\def\aminspt{$\buildrel{\prime}\over .$}

\def\about{$\sim$}
\def\msun{${\cal M}_{\odot}$}
\def\mass{${\cal M}$}

\def\x{$\times$}

\begin{document}

\title{Blue Stars and Binary Stars in NGC 6397:\\ Case Study of a Collapsed-Core 
Globular Cluster}
\author{Adrienne M. Cool and Adam S. Bolton}
\affil{San Francicso State U., 1600 Holloway Ave. San Francisco, CA 94132}

\begin{abstract}

The dense central region of NGC~6397 contains three classes of stars
whose origins are likely related to stellar interactions: blue
stragglers (BSs), cataclysmic variables (CVs), and probable helium
white dwarfs (HeWDs).  We summarize results to date concerning CVs and
HeWD candidates that have been identified in two imaging studies with
Hubble Space Telescope (HST), and present one new CV candidate that
appears well outside the cluster core.  We also present results
concerning binaries containing two main sequence stars in the central
parts of the cluster.  Proper motion information derived from two
epochs of HST data is used to remove field stars from the sample.
Binaries are then identified on the basis of their positions in the
color-magnitude diagram.  We set an upper limit of \about 3\% on the
fraction of main sequence stars with primary masses in the range
$0.45-0.8$\msun\ and mass ratios q $\ga$ 0.45.  Extrapolating to all
mass ratios gives an estimated binary fraction of $\la$ $5-7$\%.  Even
in these small numbers, such pairs are likely to be key players in the
processes that give rise to the more exotic stellar populations.

\end{abstract}


\section{Introduction}

The nearest places in the Universe where stellar collisions should be
relatively commonplace are in the central regions of globular clusters
with collapsed cores.  The nearest such cluster is NGC~6397.  At a
distance of \about $2.7\pm 0.2$ kpc (Reid 1998, Reid \& Gizis 1998,
Anthony-Twarog \& Twarog 2000), its dense central region is both
brighter and more spread out on the sky than other high-density
cluster cores.  NGC~6397 is thus a prime locus in which to study the
consequences of stellar interactions. 

The inner regions of NGC~6397 are characterized by a power-law
surface-brightness profile surrounding a resolved core with a density
of \about 10$^5$\msun\ pc$^{-3}$.  Using star counts in HST/WFPC2
images, Sosin (1997) derived a maximum-likelihood core radius of
4.8\asec (1\asecspt 5$-8$\asec\ at 95\% confidence) for stars with masses
close to the turnoff mass, consistent with values determined from the
ground (e.g., Drukier 1993).  Sosin's analysis also revealed a high
degree of mass segregation in the central regions, including a mass
function that in the inner \about 18\asec\ is inverted, i.e.,
increases with increasing mass.

The first evidence for unusual stellar populations in NGC~6397 was the
discovery of a central concentration of bright blue stragglers (BSs)
by Auri\` ere, Ortolani, \& Lauzeral (1990).  Since then,
high-resolution imaging with the Wide Field and Planetary Camera 2
(WFPC2) on Hubble Space Telescope (HST) has revealed two more classes
of stars whose origins are also likely to be linked to stellar
interactions: cataclysmic variables (CVs) and probable helium white
dwarfs (HeWDs).  The presence of CVs was first signaled by a
population of faint X-ray sources detected with ROSAT (Cool \et\
1993).  Several optical counterparts were subsequently identified via
\ha\ excess, UV excess, and/or variability (De Marchi \& Paresce 1994,
Cool \et\ 1995, 1998), and confirmed spectroscopically (Grindlay \et\
1995).  More recently, a population of blue stars similar in
brightness to CVs was found that lacked the photometric variability of
CVs (Cool \et\ 1998).  Dubbed ``non-flickerers'' (NFs), they have been
identified as probable helium white dwarfs (HeWDs) (Edmonds \et\
1999).

\begin{figure}
\plotfiddle{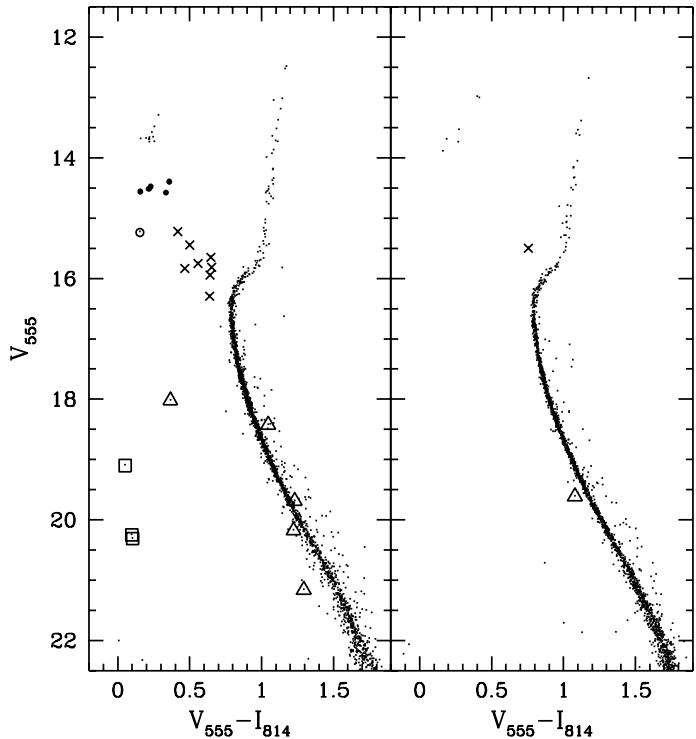}{4.0in}{0}{65}{65}{-210}{-130}
\caption{CMDs of stars within 57\asec\ of the cluster center (left)
vs.\ stars outside 57\asec\ (right).  Equal numbers of stars appear in
these two regions of the WFPC2 field.  Solid dots mark the five bright
blue stragglers in the cluster core.  The bluest of these is the one
that lies closest to the cluster center and was found by Saffer \et\
(this volume) to be most massive.  X's mark the fainter blue
stragglers, all of which lie at larger radii.  Triangles and squares
mark cataclysmic variables and probable helium white dwarfs,
respectively.}
\end{figure}

\begin{figure}
\plotfiddle{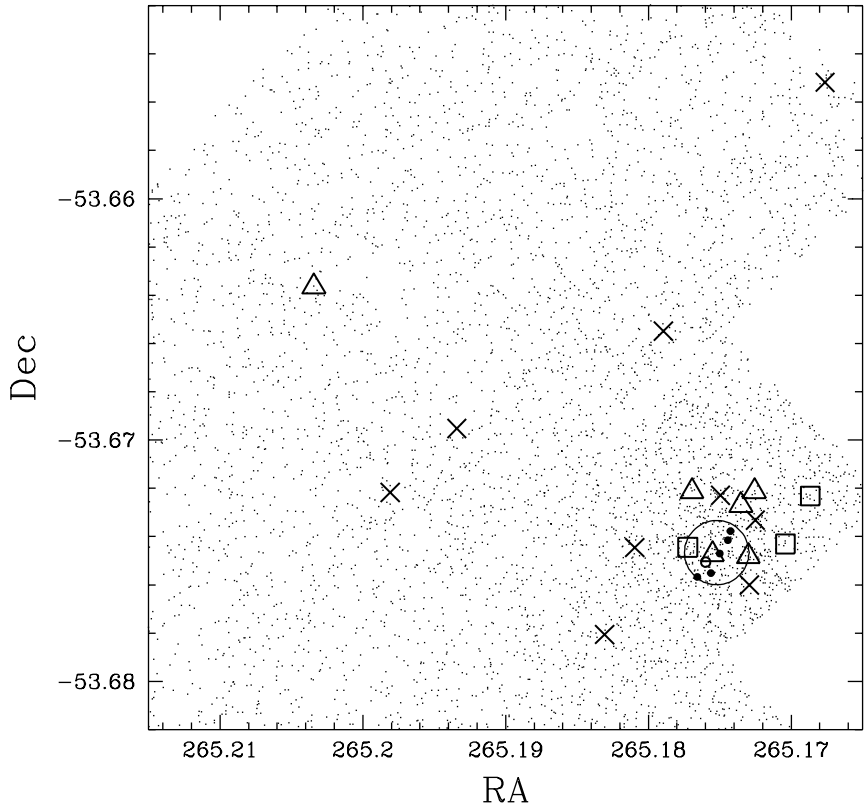}{4.5in}{0}{75}{75}{-230}{-180}
\caption{Locations within the WFPC2 field of cataclysmic variables,
helium white dwarfs candidates, and blue stragglers.  Symbols are as
in Fig.~1.  The massive blue straggler (Saffer \et, this volume) is
the one closest to the center of the cluster (middle one of the five
large dots).  The center and core radius adopted are those of Sosin
(1997): R.A. $= 17^{\rm h} 40^{\rm m} $42\secspt 06, Dec. $= -53$\deg\
40\amin\ 28\asecspt 8 (uncertainty \about 1.5\asec), and r$_c$ =
4\asecspt 8 (see text).  This center corresponds to (x,y) $=$
(529,378) on HST archive image u33r010kt (PC1 chip).  The open circle
marks a star that Saffer \et\ identify as a probable horizontal-branch
star; it had previously been identified as a blue straggler (Lauzeral
\et\ 1992).}
\end{figure}

Here we present results from two epochs of HST/WFPC2 imaging of the
central regions of NGC~6397.  Several of the results concerning CVs
and NFs have appeared elsewhere.  We collect those results and present
new color-magnitude diagrams (CMDs) showing all the CVs and NFs along
with the BSs that appear in the field.  We also describe new findings
regarding a sixth CV candidate, the first to be found many core radii
from the cluster center.  

We then discuss initial results of a study of main-sequence binaries
in the cluster.  This study makes use of the two epochs of WFPC2 data
to perform a proper-motion selection to remove field stars from the
sample of main-sequence binary candidates.  With their large cross
sections, such binaries are likely to play a critical role in stellar
interactions, affecting the dynamical evolution of the cluster as well
as the formation of exotic stellar populations.  Understanding the
populations of BSs, CVs, and NFs is likely to require an understanding
of main-sequence binaries as well.

Fig.~1 shows CMDs for all stars in the two epochs of WFPC2 data
brighter than V = 22.5.  The BSs, CVs, and NFs are all marked.  We
have divided the stars into two equally sized groups, one with r $<$
57\asec (left panel) and one with r $>$ 57\asec\ (right panel).  It
can be seen at a glance that all three populations of stars are
strongly concentrated toward the center of the cluster.  This central
concentration can also be seen in Fig.~2, in which we plot the
locations of the BSs, CVs, and NFs within the HST/WFPC2 field.

\section{Observations}

The central regions of NGC 6397 have been observed twice with the
WFPC2 camera on the Hubble Space Telescope (HST).  The first data set
(hereafter ``epoch~1'') was taken on $6-7$ March 1996, and consists
primarily of F336W (\u) and F439W (\b) images, with a small number of
short exposures in F555W and F814W.  The second data set (hereafter
``epoch~2'') was taken on $3-4$ April 1999, and consists of F656N
(\ha), F555W (\v), F675W (\r), and F814W (\i) images.  The epoch~1
observations were taken primarily for a variability study, and so were
minimally dithered; the epoch~2 observations were well dithered.

Both epochs of data were centered approximately on the cluster center,
with some adjustment made to include an off-center X-ray source in the
field.  The fields of view of the two epochs are largely, but not
completely, overlapping.  We stacked co-aligned images following
cosmic-ray removal, to produce 2 \u, 2 \b, 1 \v, and 1 \i\ image for
epoch~1, and 13 \r\, 12 \v, 12 \i, and 15 \ha\ images for epoch~2.
Hereafter we use the term ``frame'' to refer to these image stacks.

\section{Photometric and Astrometric Analysis}

Here we briefly outline the photometric and astrometric analysis
techniques; details will appear elsewhere (Bolton, Cool, \&\ Anderson
2001).  The photometric analysis was carried out using ALLFRAME
(Stetson 1994).  We constructed empirical point-spread functions for
each image, and created the star list by hand to minimize the
introduction of artifacts.  A total of 7061 stars were identified and
measured (1108, 1969, 1761, and 2223 in the PC1, WF2, WF3, and WF4
chips, respectively).

\begin{figure}
\plotfiddle{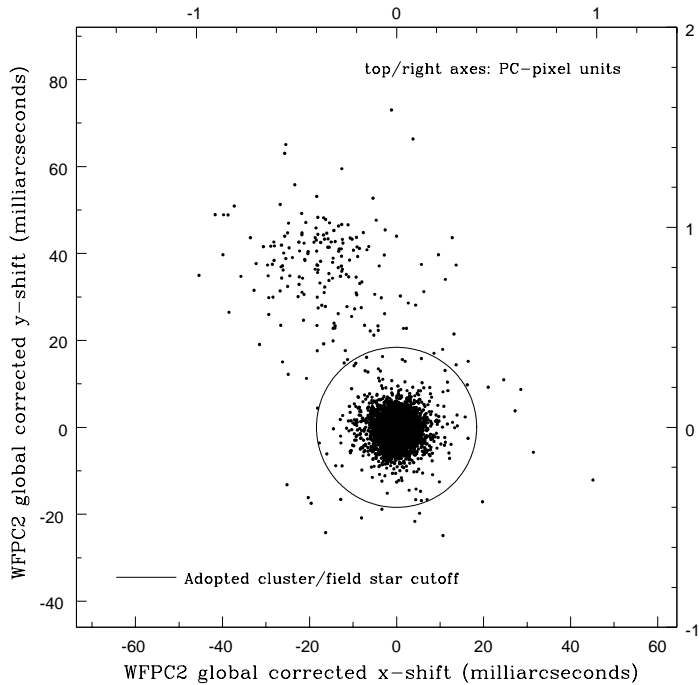}{3.5in}{0}{47}{47}{-150}{-70}
\caption{Proper motions of all stars detected in both epochs.  Because
the transformations are based primarily on cluster stars, cluster
members center around (0,0).}
\end{figure}

For the astrometric analysis, we used the technique developed by
Anderson \& King (2000) (see also Anderson, King \& Meylan 1998 and
King \et\ 1998).  We first redetermined positions of stars in each
frame in both epochs using the code developed by Anderson.  These
positions were then corrected for geometric distortion using the
method prescribed by Holtzman \et\ (1995).  We then predicted a
position in the epoch~2 images for each star in the epoch~1 images.
This was accomplished for each individual star using its 25 nearest
neighbors to determine a local general linear transformation between
the two epochs.  The chosen transformation was the one that minimized
the total square deviation between the transformed and actual
positions of the near neighbors.  Stars that yielded proper motions $>
0.4$ PC pixels (relative to the mean of cluster stars) in a first pass
were excluded from a second pass at computing the transformations
(Bolton, Cool \&\ Anderson 1999).

The differences between the observed and predicted epoch~2 positions
are shown in Fig.~3.  Cluster members center around (0,0) since the
transformations are based primarily on cluster stars (there being many
more cluster stars than field stars in the images).  A significant
population of non-members can be seen offset \about 1 PC pixel from
the cluster mean, consistent with the offset found in a similar study
of stars in another region of the cluster (King \et\ 1998).  The
separation between the two populations appears quite good.  In order
to remove the bulk of field stars from the CMDs throughout this paper,
we have eliminated stars whose displacement from the cluster mean is
more than 0.4 PC1 pixels (see circle in Fig.~3).

\section{Cataclysmic Variables}

In Fig.~4 we present results of the ALLFRAME analysis for all four
chips combined.  Three CMDs are shown side by side, with \v\ on the
vertical axis in all cases.  The colors in the left panel (\ub) come
from the epoch~1 data, while the colors in the center and right panels
(\vi\ and \hr, respectively) come from the epoch~2 data.  Only stars
that meet the criterion for membership described above are plotted.
Cataclysmic variables (CVs) are marked with triangles.  We confine our
discussion here to the brighter stars; results from a search for CV
candidates among the fainter stars will appear elsewhere (Grindlay
\et\ 2001).

\begin{figure}
\plotfiddle{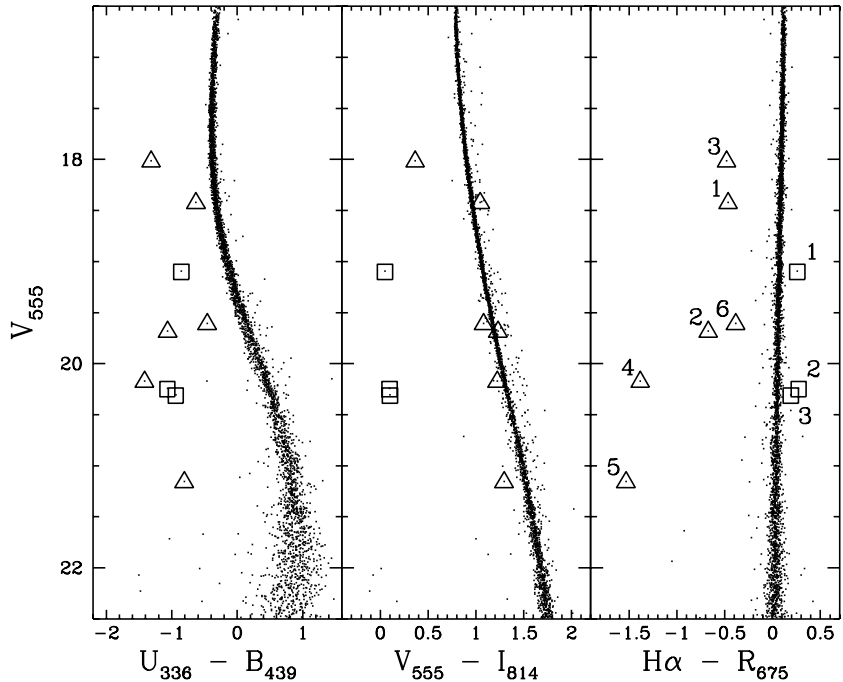}{3.5in}{270}{60}{60}{-220}{+320}
\caption{Color-magnitude diagrams for stars in all four WFPC2 chips
combined.  Cataclysmic variables are marked with triangles; helium
white dwarf candidates are marked with squares.  Numbering is explained in the
text.}
\end{figure}

CVs $1-3$ are those first identified on the basis of \ha\ excess in
Cycle~1 WFPC1 \r\ and \ha\ images (Cool \et\ 1995).  CV~4 was
identified via its variability in epoch~1 data, which also showed CVs
$1-3$ to be variables (Cool \et\ 1998).  CV~5 was identified in a
search for an optical counterpart to an X-ray source identified by
Metchev (1999) in a deep (75 ksec) ROSAT HRI exposure (see also
Verbunt \& Johnston 2000).  It appeared as a U-bright object in the
epoch~1 data, and could also just barely be seen in the wings of a
bright star in the Cycle~1 WFPC1 \ha\ images (Grindlay 1999).  These
five stars are all within \about 11\asec\ of the cluster center (see
Fig.~2), in a region of multiple overlapping ROSAT/HRI X-ray sources.
HST/FOS spectra that have been obtained of CVs $1-4$ all show Balmer
emission lines; CVs $1-3$ also show He II in emission, suggesting that
they are weakly magnetic, i.e., DQ~Her-type, systems (Grindlay \et\
1995, Edmonds \et\ 1999).

CV~6 is new.  With magnitudes and colors similar to those of other CVs
in the cluster, and with a proper motion consistent with membership
(see Fig.~5), it is a strong candidate.  It also appears to have a
faint counterpart in the deep ROSAT HRI exposure whose X-ray flux
is consistent with its being a CV (Metchev 1999; Verbunt \& Johnston
2000).  Of particular interest is its position in the cluster, roughly
72\asec, or about 15 core radii, from the center (see Fig.~2).  (Note
that it was outside the smaller field of view of the WFPC1 \ha\
study.)  It is the only one of six CVs in NGC~6397 that lies more than
about three core radii from the cluster center.

\begin{figure}
\plotfiddle{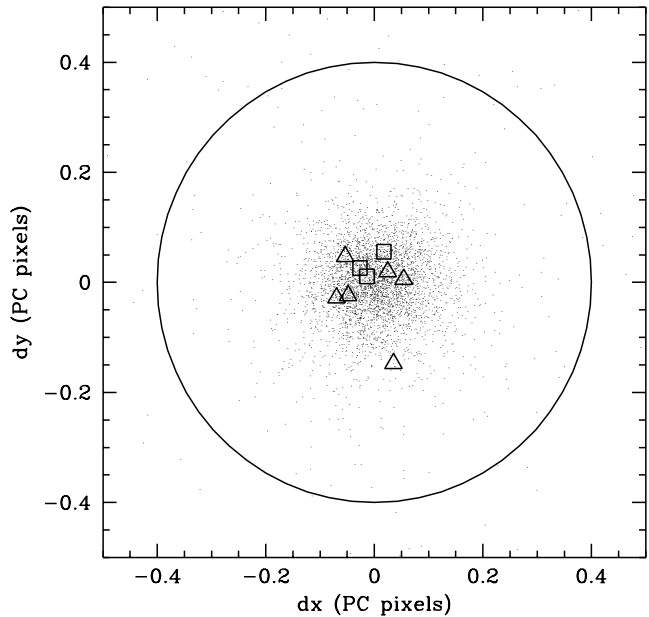}{3in}{0}{57}{57}{-190}{-150}
\caption{Close-up of a portion of Fig.~3, with CVs and NFs marked
(symbols are as in Fig.~3).  All nine stars easily meet the criterion
for proper-motion membership (see text).}
\end{figure}

\section{Helium White Dwarf Candidates}

The third class of exotic stars in the central regions of NGC~6397 was
discovered in the epoch~1 data.  Three faint blue stars were found
that distinguished themselves from CVs in two ways.  First, they
lacked the photometric variability seen in the CVs.  Second, they had
the broad-band colors of B stars rather than the UV excesses typical
of CVs.  These three stars are marked with squares in Figs. 1, 2, 4
and 5, and numbered as in Cool \et\ (1998).  The difference in the
broad-band colors of NFs vs.\ CVs can be seen by comparing the left
and center panels in Fig.~4.  In \ub\ (left panel) they appear very
similar, whereas in \vi\ (middle panel) the CVs lie close to the main
sequence while the NFs remain well to the blue side of the main
sequence.

We note that two of these ``non-flickering'' stars had been seen in
previous studies, and, being blue, were taken to be possible CVs.  NFs
2 and 3 correspond to stars 7 and 6, respectively, of Cool \et\
(1995), while NF~3 is star 4 of De Marchi \& Paresce (1994), who
identified it in HST Faint Object Camera data.

The magnitudes and colors of the NFs are consistent with those of
low-mass white dwarfs (Cool \et\ 1998).  Low-mass (helium) WDs have
larger radii than the more common \about $0.5-0.6$ \msun\
carbon-oxygen (CO) WDs.  This places them to the red side of the
sequence of $\la$ 100 CO WDs identified in the epoch~1 data (Cool,
Sosin \& King 1997), and the smaller number measured in HST images of
an off-center field in the cluster (Cool, Piotto \& King 1996).

Stronger evidence that at least one of the NFs is a helium white dwarf
was presented by Edmonds et al.\ (1999), who obtained a spectrum of NF~2
with the HST Faint Object Spectrograph.  They found that it had
a broad H$\beta$ absorption line, consistent with a high gravity of
log g $\simeq$ 6.3.  Further evidence that the other two NFs are
likely to be similar in nature to NF~2 can be seen in the right panel
of Fig.~4.  All three appear at similar offsets to the right of the
vertical main sequence in this diagram, signifying that the equivalent
widths of their \ha\ absorption lines are comparable.  More detailed
analyses of these stars are presented by Taylor \et\ (2001), who also
identify additional fainter candidate members of the class.

\section{Main-Sequence Binary Stars}

Main-sequence binaries that are compact enough to survive in a
globular cluster will be unresolved with HST at the distance of even
the nearest globulars.  But they can in principle be distinguished
photometrically, as the combined light from the two component stars
will be brighter and/or redder than that from a single main-sequence
star.  An advantage of this technique, which has been explored in
detail by Romani \& Weinberg (1991), is that it is sensitive to
binaries irrespective of their orbital period or inclination.  Thus it
is complementary to techniques that rely on photometric variability
(i.e., eclipses) or radial velocity variations (see Hut \et\ 1992 for
a review).

\begin{figure}
\plotfiddle{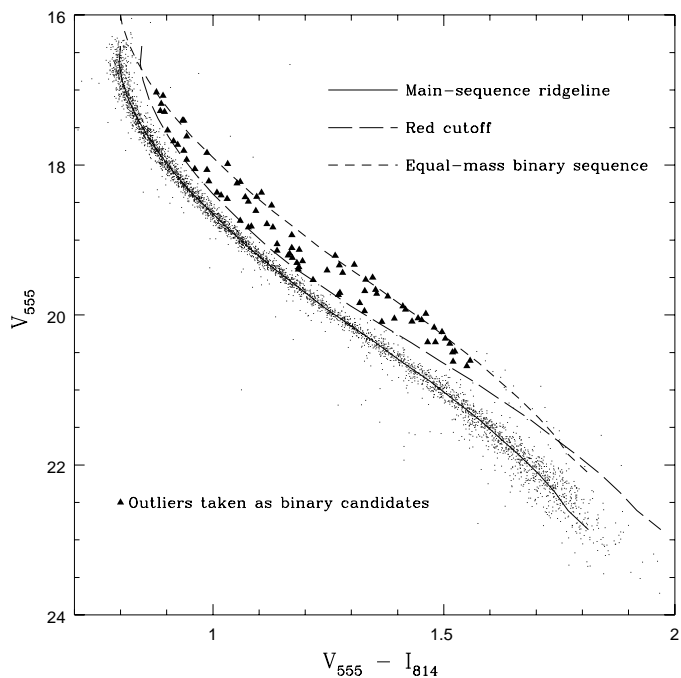}{3.2in}{0}{47}{47}{-150}{-80}
\caption{Candidate main-sequence binary stars in the WFPC2 field.}
\end{figure}

\begin{figure}
\plotfiddle{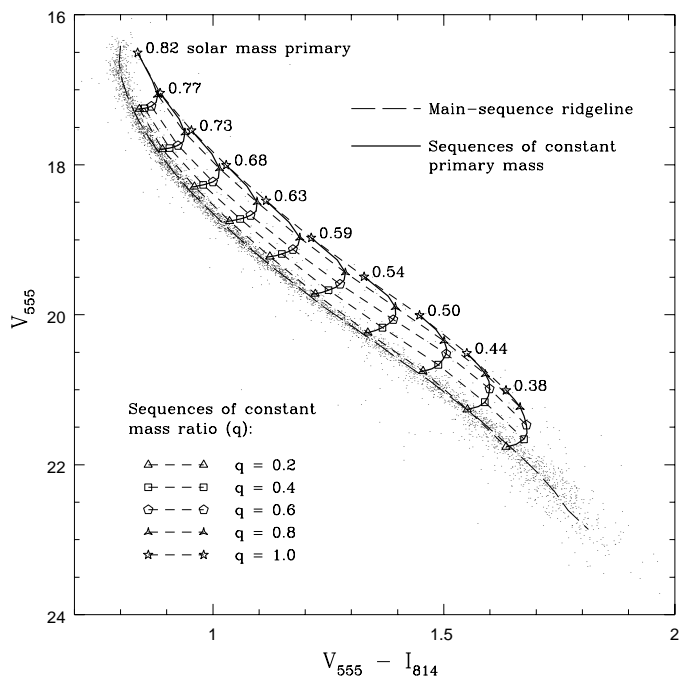}{3.2in}{0}{47}{47}{-150}{-80}
\caption{Model binary sequences constructed using our empirical
main-sequence ridge line and the model mass--luminosity relations of
Alexander \et\ (1997).  A distance modulus of $(M-m)_V = 12.6$ and a
reddening of $E(V-I) = 0.19$ were adopted for this study.}
\end{figure}

In Fig.~6 we show the CMD resulting from our ALLFRAME analysis of the
WFPC2 \v\ and \i\ images.  Only proper-motion members are plotted.  To
select candidate binaries, we began by fitting a ridge line to the
main sequence.  This was accomplished by determining the mean \vi\
color in 0.25 magnitude bins along the sequence, with outliers
excluded, and iterating until the mean converged for each bin.  Within
each iteration, the boundary between main-sequence stars and
``outliers'' was selected so that over the whole sequence fewer than
0.5 stars would be expected to lie beyond the boundary if the errors
in color were Gaussian.  The final boundary determined in this way
appears as the ``red cutoff'' in Fig.~6.  We identify stars redward of
this boundary as binary candidates.  Some portion of them will be
chance superpositions of unrelated stars in the cluster, an issue to
which we will return below.

\begin{figure}
\plotfiddle{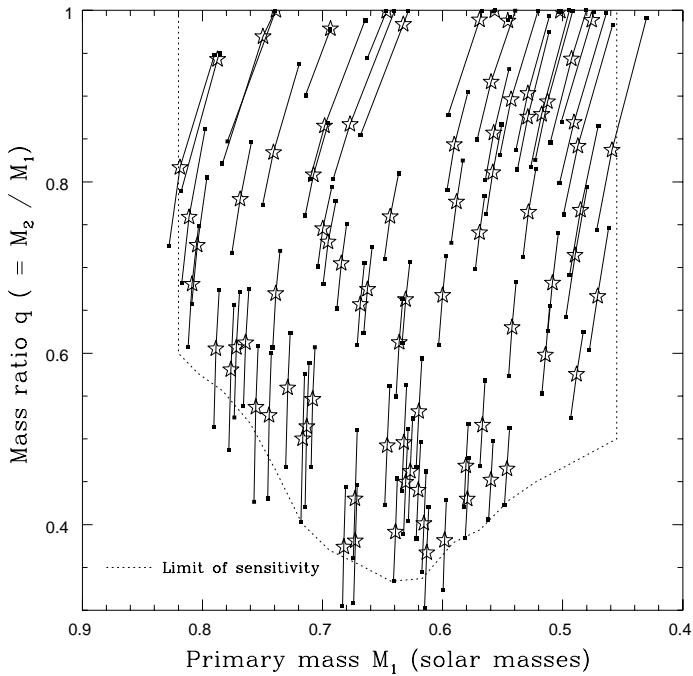}{3.5in}{0}{47}{47}{-150}{-70}
\caption{Primary mass vs.\ mass ratio for candidates shown in Fig.~7.
Photometric uncertainties translate into highly eccentric error
ellipses, the major axes of which are plotted.  Uncertainties in the
mass ratio are typically at the level of $\la$ 0.1 and in some cases
are as low as \about 0.05.}
\end{figure}

A total of 81 stars appear in the binary region of Fig.~6.  This
represents \about 3\% of the \about 2500 stars in the magnitude range
\v\ \about $17-21$.  This figure is not representative of the entire
population of main-sequence binaries, however.  Some will be missed,
namely those for which the secondary star contributes too little light
to the system to shift its color appreciably off the main-sequence
ridge line.

The limitation imposed by mass ratio is illustrated in Fig.~7.  As
increasingly massive secondary stars are combined with a primary star,
the binary's light first becomes redder, then brighter, then turns
back to the blue until, for equal mass systems, the combined light is
the same color but twice as bright as that of the primary star.  In
order to identify specific stars as candidate binaries (as opposed to
doing a statistical analysis--see, e.g., Romani \& Weinberg (1991)),
sufficiently large mass ratios are required in order to pull the star
far enough away from the ridge line of the main sequence that it can
be distinguished from a single star.

\begin{figure}
\plotfiddle{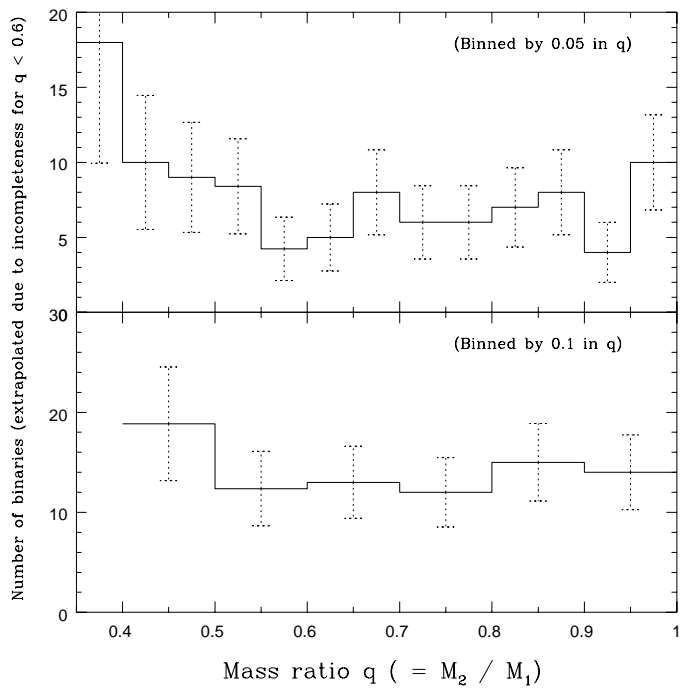}{3.5in}{0}{47}{47}{-150}{-70}
\caption{Histogram of mass ratios of binary candidates.  Error bars
are based on root-N statistics.}
\end{figure}

We have determined best values for the primary mass and mass ratio of
each of the candidate binaries by comparing their \v\ and \i\
magnitudes to those of a grid of model binaries like those shown in
Fig.~7.  The results are shown in Fig.~8.  Here it can be seen that we
are sensitive to binaries with primary masses in the range \mass $_1$
\about $0.8-0.45$\msun\ and mass ratios as low as \about 0.4.  A more
typical limiting mass ratio is q $\ga$ 0.45.  This sensitivity range
is set by a combination of the slope of the main sequence and the
photometric error as a function of magnitude.  Based on these
sensitivity limits, we place an upper limit of 3\% on the binary
fraction among main sequence stars with primary mass in the range
$0.45-0.8$ \msun\ and mass ratios q $\ga$ 0.45.  

This figure represents an upper limit because a portion of the
candidates we identify will be chance superpositions of two physically
unrelated stars in the cluster.  The WFPC2 images are not tremendously
crowded, with \about $1000-2000$ stars per 800\x 800-pixel chip, owing
in part to the relative proximity of NGC~6397.  But with the binary
fraction being so low, chance coincidences are a potentially important
factor.  We estimate that as many as $1 \over 3$ to $2 \over 3$ of the
candidates could be chance superpositions.  Non-Gaussian errors could
also push more stars into the part of the CMD associated with
binaries.  Additional work is underway to try to constrain the numbers
of chance coincidences more precisely.  If a significant portion of
the candidates are superpositions, the binary fraction in the cluster
will be correspondingly reduced.  It could also impact our preliminary
conclusions about the distribution of binaries as a function of
primary mass and mass ratio.  Nevertheless, the 3\% figure we have
determined represents a firm upper limit on the main-sequence binary
fraction for q $\ga$ 0.45.

We can take this one step further to obtain an estimate of the total
main-sequence binary fraction by extrapolating to lower mass ratios.
Judging from Fig.~8, binaries are fairly uniformly distributed in
\mass $_1$ and q.  Binning by mass ratio produces the results shown in
Fig.~9, which also suggest that the distribution of binaries as a
function of q is quite flat.  Extrapolating this flat distribution to
lower mass ratios, we estimate the total main-sequence binary fraction
in the central regions of NGC~6397 to be $\la$ $5-7$\%.

\section{Discussion}

Stars in the middle of dense clusters must collide.  Newton's laws and
the laws of probability require it.  But the outcomes of the
collisions, and especially of the more numerous near misses, are less
certain.  With HST it is possible to observe the products of these
interactions directly.  In NGC~6397, every type of star seems to be in
on the action.  Main sequence stars are combining to form blue
stragglers; white dwarfs are coupling with main-sequence stars to form
cataclysmic variables; giants are being stripped of their envelopes to
reveal their helium cores.

The details of the particular processes that are generating this
stellar mena\-gerie will take time to sort out.  Multiple pathways
exist by which each of these classes of stars may form, and the
presence of binaries, even in the small numbers found here, multiplies
the possibilities.  Formation rates, lifetimes, and
destruction/ejection rates all need to be factored in to make sense of
the variety and distributions of objects being found.  The growing
sophistication of simulations of clusters and stellar interactions,
coupled with improved constraints on the numbers and distributions of
both ordinary and extraordinary stars in clusters, holds out the
promise of significant progress in the near future.  New dynamical
modeling of NGC~6397, making full use of the new information HST has
provided, would clearly be valuable.  Here we draw attention to a few
points of particular interest that may bear further investigation.

\vskip 0.05in

(1) Why are there so few main-sequence binaries in NGC~6397?

\vskip 0.05in

The limit that we place on the main-sequence binary fraction in
NGC~6397 is one of the lowest reported for a globular cluster.  Most
values, obtained using a variety of different techniques, range upward
of 10\% (see Hut \et\ 1992 for a review).  Particularly notable is the
much higher fraction (15\%--38\%) measured by Rubenstein \& Bailyn
(1997) in the center of NGC~6752, another nearby, core-collapsed
cluster with otherwise similar properties.

Perhaps NGC~6397 is farther along in the process of burning up its
original store of binaries than NGC~6752.  The short relaxation time
in NGC~6397 is likely to facilitate the rapid feeding of binaries into
the core where they will quickly be modified through interactions with
other stars (McMillan \& Hut 1994, and references therein), perhaps
even to form some of the blue stragglers now seen in the core.  But
the present dynamical properties of NGC~6752 are not so different from
those of NGC~6397 as to make this a particularly satisfying
hypothesis.  Whether their past histories (e.g., the timing of core
collapse) might be sufficiently different one can only speculate
about, and the possibility remains that the two clusters were simply
born with significantly different binary populations.

Still, it appears likely that the main-sequence binaries we now
observe in NGC~6397 are a vestige of the original population.  Large
interaction cross sections coupled with the short relaxation times in
NGC~6397 (\about 10$^5$ yr in the center; \about $2\times 10^8$ yr at
the half-mass radius of 2\aminspt 8---Djorgovski 1993) make binaries extremely
vulnerable to destruction through a variety of direct and indirect
processes.  Even if the binaries we are now seeing were formed when
the cluster was born, it appears that few will been left unscathed.
Judging from the calculations of Davies (1995), the rates of binary
hardening and exchange interactions should be sufficiently high in
this cluster that most systems will have significantly reduced
periods, and some fraction of their present components will not be
original.  Thus it remains to be seen, for example, whether the
distribution of mass ratios that we derive has more to say about the
cluster's primordial population of binaries or about the nature of
exchange interactions.  Either way, the fraction of these binaries,
while small, is large enough to contribute significantly to the
production of CVs, NFs, and/or BSs (Davies 1995).

\vskip 0.05in

(2)  What accounts for the distribution of BSs, CVs, and NFs?

\vskip 0.05in

Saffer \et\ (this volume) have measured masses for the five bright
blue stragglers in the core of NGC~6397 (see Fig.~2).  Four have
masses of about twice the turnoff mass; the fifth is of order three
times the turnoff mass.  Interestingly, the latter is the blue
straggler closest to the cluster center---exactly where one might
expect the most massive objects to reside (see Fig.~2).  Its position
is consistent with being at the cluster center, according to the star
count analysis of Sosin (1997) as well as the ground-based analysis of
Auri\` ere \et\ (1990).

The distribution of the remaining bright BSs is more puzzling.  Why
are they so much more concentrated toward the center than the CVs,
which ought to have comparable (though somewhat lower) masses?  The
NFs present a puzzle of their own.  Within the limits of small-number
statistics, their distribution appears similar to that of the CVs.
Yet if they are He WDs, their masses are much too low (\about
0.25\msun) to account for this similarity.  A natural solution is that
they have binary companions.  Companions are needed in any case to
explain the formation of a He WD, whether through Roche-lobe overflow
or via common-envelope evolution following the collision of a red
giant with another star.  Sufficiently massive main-sequence
companions are ruled out (Cool \et\ 1998), but neutron stars, or
possibly massive white dwarfs, are viable.  Edmonds \et\ (1999) find
preliminary evidence for binarity in the radial velocity of the one NF
for which a spectrum has been obtained; additional studies are clearly
needed.

This still leaves open the question of why the bright BSs are much
more concentrated than either the CVs or NFs.  One factor that may
distinguish the BSs from the CVs and NFs is binarity.  The CVs are
certainly binaries; the NFs make little sense as He WDs unless they
too are binaries.  The bright BSs show no evidence for photometric
variability that would suggest binarity (in contrast to some of the
fainter BSs in the cluster---Cool \et\ 2001).  Perhaps the recoil that
binaries experience in interactions with other stars keeps them more
spread out, on average, than single stars of comparable mass.

A further puzzle is that all three populations (BSs, CVs, and NFs)
appear much more centrally concentrated than the population of
1.4\msun\ neutron stars that Dull (1996) included in his Fokker-Planck
model of NGC~6397.  Yet their masses should all be roughly comparable.
This suggests that not all of these objects have come into equilibrium
with the other stars in the cluster, despite having lifetimes well in
excess of the relevant 2-body relaxation times.  Meanwhile, Sosin's
(1997) analysis of the epoch~1 HST data suggests that the
main-sequence stars {\it are} in energy equipartition with one another
within the cusp (radius \about 100\asec).  It will be interesting to
see whether this apparent discrepancy between the behavior of
main-sequence stars and probable interaction products is borne out in
future dynamical model that take account of the HST results.

\vskip 0.05in

(3) Why is CV~6 out at \about 15 core radii?

\vskip 0.05in

Numerical simulations have shown that close binary stars in dense
cluster cores are vulnerable to interactions with passing stars that
can eject the binaries from the cluster.  If its recoil velocity is
insufficient to escape from the cluster, the binary may spend
considerable time at large radii before sinking back to the central
region of the cluster.  If this has happened to CV~6, it appears that
it would likely have been a relatively recent occurrence.  The central
relaxation time in NGC~6397 is exceedingly short; even at the
half-mass radius (\about 170\asec) it is only \about 2\x 10$^8$ yr.
Unless its radial distance from the center is much greater than its
72\asec\ projected distance, CV~6 shouldn't take long to migrate back
into the center.

In view of the possibility that CV~6 has been ejected from the center
in the relatively recent past, it appears somewhat tantalizing at
first glance that in the proper motion diagram of Fig.~5, CV~6 is the
triangle offset from the rest of the CVs and NFs.  However, CV~6 is
the only CV that lands on a WF chip rather than on the PC1 chip.  The
larger WF pixels will result in a correspondingly larger uncertainty
in the measurement of its proper motion.  Moreover, the direction of
motion implied by the proper motion (assuming for the moment that the
offset from (0,0) is significant) is closer to being tangential than
radial.  We conclude that it is most likely that the offset of the
proper motion of CV~6 from (0,0) is little more than measurement
error.  Further refinements by Anderson and King of the astrometric
techniques they have developed for WFPC2 may permit direct
measurements of the internal motions of cluster stars in the near
future (see, e.g., Anderson \et\ 1998).  Such measurements of this and
the other CVs, NFs, and BSs in NGC~6397 would be of particular
interest.

\acknowledgments

We are grateful to our collaborators for help and discussions at many
junctures in this project: J. Anderson, J. Grindlay, P. Edmonds,
J. Taylor, C. Sosin, I. King, P. Lugger, H. Cohn, and C. Bailyn.  We
particularly wish to thank Jay Anderson for his help with the
astrometric analysis and generosity with other software.

\end{document}